\def\rfr#1{eq. (\ref{#1})}

\def\virg#1{``#1''}

\def\eqi{\begin{equation}}
\def\eqf{\end{equation}}
\def\eqia{\begin{eqnarray}}
\def\eqfa{\end{eqnarray}}
\def\rp#1#2{{#1\over#2}} \def\lb#1{\label{#1}}

\def\Om{\mathit{\Omega}}
\def\kap{\bds{\hat{k}}}
\def\kx{\hat{k}_X}
\def\ky{\hat{k}_Y}
\def\kz{\hat{k}_Z}

\def\ci{\cos I}

\def\sO{\sin\Om}
\def\cO{\cos\Om}
\def\bds#1{\boldsymbol{#1}}

\documentclass[doublecol]{epl2}

\title{Some considerations on the present-day \textcolor{black}{results for the detection of} frame-dragging after the final \textcolor{black}{outcome} of GP-B}
\shorttitle{Some considerations on the present-day results for the detection of frame-dragging etc.} 

\author{L. Iorio\inst{1,2,3} }
\shortauthor{L. Iorio}

\institute{
  \inst{1} Ministero dell'Istruzione, dell'Universit\`{a} e della Ricerca (MIUR)-Istruzione\\
  \inst{2} International Institute for Theoretical Physics and
High Mathematics Einstein-Galilei (IITPM)\\
\inst{3} Fellow of the Royal Astronomical Society (FRAS)
}
\pacs{04.80.-y}{Experimental studies of gravity}
\pacs{04.80.Cc}{Experimental tests of gravitational theories}
\pacs{91.10.Sp}{Satellite orbits}

\abstract{
The cancelation of the first even zonal harmonic coefficient $J_2$  from the linear combination $f^{\rm (2L)}$ of the nodes $\Om$ of LAGEOS and LAGEOS II used in the latest tests of the  Lense-Thirring effect cannot be perfect, \textcolor{black}{contrary to what} assumed so far. It is so \textcolor{black}{also} because of the uncertainties in the spatial orientation of the terrestrial spin axis $\bds{\hat{k}}$. As a consequence \textcolor{black}{of above}, the coefficient $c_1$ entering $f^{\rm (2L)}$, which is not a solve-for parameter being, instead, theoretically computed from the analytical expressions of the classical node precessions $\dot\Om_{J_2}$ due to $J_2$, is, on average, uncertain at a $10^{-8}$ level over multi-decadal time spans $\Delta T$ comparable to those used in the data analyses performed so far. A further $\simeq 20\%$ systematic uncertainty, thus, occurs. The shift $\Delta\rho_{\rm LT}$ due to the gravitomagnetic frame-dragging on the station-spacecraft range $\rho$ is numerically computed over $\Delta T=15$ d and $\Delta T=1$ yr. The need \textcolor{black}{to look}  at such a directly observable quantity is \textcolor{black}{highlighted}, along with some critical remarks concerning the methodology used so far to measure the Lense-Thirring effect with the LAGEOS satellites. Suggestions for a different, more trustable and reliable approach are offered.}

\begin{document}

\maketitle
\section{Introduction}
In the Einsteinian general relativity, which is \textcolor{black}{a} fully Lorentz-invariant \textcolor{black}{theory of gravitation}, matter-energy currents create an additional, magnetic-like \textcolor{black}{component of the} gravitational field \cite{gvm} with respect to the static case. It is believed to play a relevant role in explaining relativistic jets ejected from active galactic nuclei \cite{thorne,stella}. The gravitomagnetic field of a  rotating body affects orbiting test particles, precessing gyroscopes,
moving clocks and atoms, and propagating electromagnetic
waves with a variety of phenomena \cite{scia}. \textcolor{black}{Some of them have been  put to the test more or less recently}. For an overview of such a phenomenology  in the solar system, see, e.g., \cite{review}.

The Gravity Probe B (GP-B) experiment \cite{mission} officially came  to an end, with the release of its final results \cite{GPB} according to which the general relativistic gravitomagnetic gyroscope precession \cite{gyro1,gyro2} would have  been  measured with a claimed accuracy of $19\%$. \textcolor{black}{Such a figure is \textcolor{black}{greater} than the previously expected $1\%$ level because of a number of unwanted systematic errors whose proper treatment required much additional efforts by the GP-B team \cite{analisigpb}. Independent analyses  by different teams will be important in critically assessing the reliability of the results  of \cite{GPB}. This is beyond the scope of the present paper. It} is now even more important than before to critically scrutinize the competing tests of the gravitomagnetic Lense-Thirring orbital precessions \cite{LT} performed in the past years \cite{ciu09} with the LAGEOS  and LAGEOS II SLR satellites in the gravitational field of the Earth, as originally proposed in \cite{cugu}. Let us recall that the GP-B mission was a dedicated experiment in the terrestrial gravitational field costed US$\$$ 750 million  and lasted 52 yr, while the \textcolor{black}{gravitomagentic} data analyses \cite{analisi} of the  LAGEOS spacecrafts, \textcolor{black}{which were originally launched for different purposes},  were much less expensive and comparatively less extended in time. According to I. Ciufolini, its
accuracy would be $10\%$ or better \cite{ciu04}; for  recent articles establishing a comparison between GP-B and the previous LAGEOS-based results, see \cite{nature,blog1,blog2} in which it is basically argued that GP-B would have just reached the same results of the earlier tests with the LAGEOS spacecraft at a much higher cost and with an even worst, or, at most, comparable, accuracy.
\section{Can the cancelation of the effect of the quadrupole mass moment of the Earth in the LAGEOS-based tests be perfect?}
The following linear combination of the longitudes of the ascending nodes\footnote{The longitude of the ascending node $\Om$ is one of the angles determining the orientation in space of the satellite's Keplerian ellipse.} $\Om$ of   LAGEOS and LAGEOS II   \cite{ries,iorio03}
\eqi f^{(\rm 2L)}\doteq \Om^{\rm (L)}+c_1\Om^{\rm (L\ II)}\lb{combo}\eqf was used
in the tests of the Lense-Thirring effect performed so far  with such artificial bodies \textcolor{black}{orbiting} the Earth. Frame-dragging was purposely not modeled \cite{ciu04}, and time series of\footnote{The term \virg{residual} is, actually, improper for the node. Indeed, all the Keplerian orbital elements are not observable quantities. They can only be computed at various epochs from the corresponding state vectors in cartesian coordinates which, in turn, are computed from the measured values of the direct observables.} \virg{residuals} of the nodes \cite{luc1,luc2} of both satellites, combined according to \rfr{combo}, were analyzed and subsequently fitted with a straight line plus other time-dependent signals.
The coefficient $c_1$ entering \rfr{combo} is not one of the \textcolor{black}{several} solve-for parameters estimated in the data reduction process. Following an approach set forth  in a different context \cite{ciu96}, its value is theoretically computed
as \cite{iorio03}
\eqi c_1\doteq-\rp{\dot\Om_{J_2}^{\rm (L)}}{\dot\Om_{J_2}^{\rm (L\ II)}}\lb{coeff}\eqf
from the analytical expressions of the classical secular node precessions $\dot\Om_{J_2}$
 of both the LAGEOS satellites caused by the first even zonal harmonic coefficient $J_2$ of the expansion in multipoles of the Newtonian part \textcolor{black}{$U_{\rm N}$} of the terrestrial gravitational potential. \textcolor{black}{This multipolar expansion of $U_{\rm N}$} accounts for its departure from spherical symmetry because of the centrifugal deformation due to the Earth's diurnal rotation \cite{Cap}.
 Traditionally, \rfr{coeff} has always been computed \cite{iorio03} so far \textcolor{black}{from} the well known expression \cite{Cap}
 \eqi\dot\Om_{J_2}=-\rp{3nJ_2 R^2\cos I}{2a^2\left(1-e^2\right)^2}\lb{precej2}.\eqf
 In \rfr{precej2} $a$ is the semi-major axis of the satellite's orbit, $e$ is its eccentricity, $I$ is its inclination to the reference $\{X,Y\}$ plane, assumed to be coincident with the Earth's equator, $R$ is the terrestrial equatorial radius, and $n\doteq\sqrt{GM/a^3}$ is the Keplerian mean motion of the satellite with respect to the Earth whose mass is denoted with $M$; $G$ is the Newtonian constant of gravitation. The orbital parameters of LAGEOS and LAGEOS II, referred to a geocentric inertial system, are shown in table \ref{tavola}.
\begin{table*}
\caption{ Keplerian orbital parameters of LAGEOS and LAGEOS II computed from state vectors, in cartesian coordinates, corresponding to a given epoch kindly provided by L. Combrinck to the author. The inclination $I$ and the node $\Om$ refer to a geocentric inertial system whose reference $\{X,Y\}$ plane is assumed to be coincident with the  Earth's equator. The semi-major axes and the angles are given with a cm-level and mas-level accuracy, respectively ($1$ cm$=10^{-5}$ km, $1$ mas$=2.7\times 10^{-7}$ deg).  }
\label{tavola}
\begin{center}
\begin{tabular}{lllll}
Spacecraft  & $a$ (km) & $e$ & $I$ (deg) & $\Om$ (deg) \\
LAGEOS & $12274.75303$ & $0.0039962$ & $109.8617388$ & $-1.4477848$ \\
LAGEOS II & $12159.19724$ & $0.0141892$ & $52.6013013$ & $-94.7543331$
\end{tabular}
\end{center}
\end{table*}

 The aim of \rfr{combo}, with \rfr{coeff}, is to cancel out, by construction, such precessions. \textcolor{black}{Since they are} nominally 7 orders of magnitude \textcolor{black}{greater} than the Lense-Thirring ones
 \eqi \dot\Om_{\rm LT}=\rp{2GS}{c^2 a^3(1-e^2)^{3/2}},\lb{ltprece}\eqf where $S$ is the Earth's angular momentum and $c$ is the speed of light in vacuum, \textcolor{black}{they} represent  a major source of systematic bias in determining them. For the LAGEOS satellites \rfr{ltprece} yields about 30 milliarcseconds per year (mas yr$^{-1}$ in the following), so that the combined Lense-Thirring signal amounts to approximately 50 mas yr$^{-1}$ according to \rfr{combo}. In principle, such a removal of $J_2$ from \rfr{combo} is exact, or \textcolor{black}{so} it has always been \textcolor{black}{considered}  until now. Indeed, in all the more or less realistic evaluations of the systematic error due to the geopotential existing in literature \cite{ciu09,analisi,review}, the part due to $J_2$ was always set to zero by definition and independently of $\sigma_{J_2}$. \textcolor{black}{Thus, the focus was} on the impact of the other, uncanceled even zonal harmonics of higher degree $J_{\ell}, \ell=4,6,8,\ldots$, \textcolor{black}{known} with a certain level of \textcolor{black}{uncertainty}. Actually, the effect of $J_2$ on \rfr{combo} cannot be exactly zero because of a number of factors.

 One of them relies on the fact that, for a given set of values\footnote{They were never explicitly specified in the analyses performed so far, by assuming for them some standard figures \cite{ciu09}, approximately representative of the orbital configurations of the LAGEOS satellites.} of the satellites' orbital parameters from which $c_1$ is computed by means of \rfr{coeff} and \rfr{precej2}, the actual accuracy with which $c_1$ can be known is necessarily limited by the uncertainties with which the satellites' Keplerian orbital elements of interest can be determined in the data reduction procedure. It was recently shown \cite{grg} that $\sigma_a\simeq 2$ cm and $\sigma_I\simeq 0.5$ mas  yield $\Delta c_1\simeq 10^{-8}$,  corresponding to a further systematic uncertainty of about $20\%$ in the Lense-Thirring signature. If, instead, one optimistically assumes $\sigma_a\simeq 2$ cm and \cite{ciu09} $\sigma_I\simeq 10-30$ $\mu$as, then $\Delta c_1\simeq 8\times 10^{-9}$, which implies an additional $14\%$ bias. On the contrary, $c_1$ was always \textcolor{black}{released} so far with a very limited number of significant digits; for example, in \cite{ciu09} we have $c_1=0.545$.  \textcolor{black}{As pointed out in \cite{grg}, it would be incorrect to argue that the impact of $\Delta c_1$ would be negligible since it should be multiplied by the uncertainty in $J_2$. Indeed, the standard error propagation theory tells us that, in addition to the mixed, cross-correlated terms containing the products of the uncertainties, there are also the linear terms proportional to the uncertainties in each parameter. Moreover, in the LAGEOS tests both $c_1$ and $J_2$ are not estimated solve-for parameters}. For the sake of definiteness, we will denote the values of $c_1$ obtained from \rfr{precej2} with $c_1^{(0)}$;  table \ref{tavola} yields $c_1^{(0)}=0.540976405$.

\begin{figure}
\onefigure[width=8 cm]{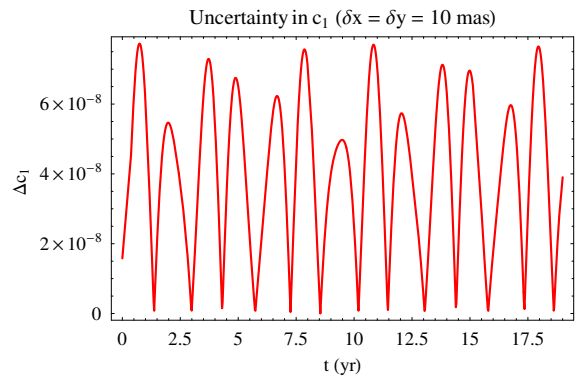}
\caption{Temporal evolution of the difference $\Delta c_1$ between the value of $c_1^{(0)}$ computed by assuming the Earth's spin axis exactly coincident with the reference $Z$ axis of an inertial equatorial reference system, and the value of $c_1^{(\sigma_{\bds{\hat{k}}})}$ computed by assuming an uncertainty $\sigma_{\bds{\hat{k}}}$ of the order of 10 mas in the orientation $\bds{\hat{k}}$ of the Earth's spin axis in the same reference system. The time span is $\Delta T= 19$ yr, while the time step is $\Delta t=7$ d. The initial conditions chosen for LAGEOS and LAGEOS II are those listed in table \ref{tavola}. $\Delta c_1$ is characterized by an average $\left\langle\Delta c_1\right\rangle=4.1\times 10^{-8}$, and a peak-to-peak amplitude of $\Delta c_1^{(\rm max)}-\Delta c_1^{(\rm min)}=7.7\times 10^{-8}$. }
\label{figura}
\end{figure}

 Another issue, not yet considered in literature, is that it is incorrect to assume a perfect alignment of the Earth's spin axis, whose unit vector is denoted by $\bds{\hat{k}}$, and the reference $Z$ axis of the \textcolor{black}{geocentric inertial} system actually used. Indeed, on the one hand, the latter refers to a given reference epoch, typically J2000.0, while the time spans $\Delta T$ over which the data of LAGEOS and LAGEOS II were analyzed necessarily cover 19 yr or less: during such a temporal interval $\kap$ did not remain fixed in the inertial space due to a variety of physical processes \cite{libro}. Such changes, even if taken into account and  modeled, are, of course, known only  with a limited accuracy \cite{Seidel}. On the other hand, it is well known that another source of uncertainty in the location of $\kap$ is given by the polar motion \cite{libro} with respect to the Earth's crust itself, known with an accuracy of about  $10-20$ mas \cite{libro,IERS} over a time interval of just 1 yr. See also http://www.iers.org/nn$\_$10398/IERS/EN/Science/\\
 EarthRotation/PolarMotion.html?$\_\_$nnn=true. Thus, it is important to quantitatively assess the further systematic error $\Delta c_1$ induced by the use of $c_1^{(0)}$ with respect to values, denoted as $c_1^{(\sigma_{\kap})}$,  computed by taking into account the real spatial orientation of $\bds{\hat{k}}$.
To this aim, a first step consists of  computing the long-term node variations $\dot\Om_{J_2}$ for a generic orientation of $\kap$. The  acceleration experienced by a test body orbiting an oblate central mass rotating about a generic direction $\kap$ is \cite{libro}
\eqi\bds A_{J_2}=-\rp{3GM J_2 R^2 }{2r^4}\left\{\left[1-5\left(\bds{\hat{r}}\bds\cdot\kap\right)^2\right]\bds{\hat{r}}+2\left(\bds{\hat{r}}\bds\cdot\kap\right)\kap\right\}.\lb{accel}\eqf
Since its magnitude is quite smaller than the main Newtonian monopole, its effect on the  particle's orbital motion can be straightforwardly worked out with standard perturbative techniques. The Gauss equation for the variation of the node \cite{befa} allows to obtain the rate of change of $\Om$ averaged over one orbital revolution. It turns out to be
\eqi
\dot\Om_{J_2} = \rp{3n J_2 R^2}{4 a^2\left(1-e^2\right)^2}\mathcal{F}(I,\Om;\bds{\hat{k}})\lb{prece},\eqf
with
\eqi
\begin{array}{lll}
\mathcal{F}&\doteq & 2\kz\cos 2I\csc I\left(\kx\sO-\ky\cO\right)+ \\ \\
&+&\ci\left[\kx^2+\ky^2-2\kz^2+\left(\ky^2-\kx^2\right)\cos 2\Om-\right. \\ \\
&-&\left. 2\kx\ky\sin2\Om\right].\lb{effe}
\end{array}
\eqf
It is an exact result in $e$ and $I$ in the sense that no a-priori simplifying assumptions on their values were assumed; in general, it can also be useful in other contexts involving different central bodies and test particles \cite{review}.
It can be noticed that, according to \rfr{prece} and \rfr{effe},  the long-term rate of change of $\Om$ consists of the sum of a genuine secular precession and of a harmonic, time-dependent signal involving $\Om$ and $2\Om$.
Moreover, \rfr{effe} reduces to
\eqi \mathcal{F}=-2\ci\eqf for $\kx=\ky=0,\kz=\pm 1$, yielding the well-known secular precession of \rfr{precej2}.
We will denote the value of $c_1$ computed from \rfr{prece}-\rfr{effe} by $c_1^{(\sigma_{\kap})}$. In Figure \ref{figura} we plot the uncertainty in $c_1$ raising from having used just $c_1^{(0)}$ over a temporal interval $\Delta T=19$ yr representative of the time spans actually used in real data analyses, and for a 10 mas uncertainty in the position of $\kap$.
It can be noticed that its impact is non-negligible since it is of the order of $4-8\times 10^{-8}$, implying a further $\simeq 20\%$ systematic uncertainty in the gravitomagnetic signature.
\section{What was  really measured in the LAGEOS-based tests?}
In SLR studies, the directly observable quantity is the range $\rho$ between a spacecraft equipped with retroreflectors and a ground-based station\footnote{It is just the case to recall that ranges refer to an Earth-fixed rotating reference system. In order to obtain the values of table \ref{tavola} one has to take into account the polar motion, the Earth rotation, the precession and the nutation. See \cite{libro} for details.} \cite{libro}. It is straightforwardly computed by multiplying  $c$ by the time interval elapsed between the emission of the laser pulse sent to the orbiting target body and its subsequent reception after it was bounced back by the retroreflectors onboard the satellite. The precision  of such measurements is nowadays at the mm level \cite{libro}. Post-fit range residuals for good targets like LAGEOS and LAGEOS II, obtained after the adjustment of a number of solved-for parameters pertaining to the satellites' physical properties and orbital dynamics, and the measurement process itself, are as large as 1 cm or less in a Root-Mean-Square (RMS) sense \cite{libro}. They globally reflect the impact of all the unmodeled and mismodeled sources of errors like, e.g., some unknown or poorly modeled forces acting on the satellites. The post-fit range residuals are also a measure  of the effectiveness of the orbit determination process in which the estimated values of some parameters may partly or totally absorb the effects of other parameters not included in the list of those to be adjusted, or of totally unmodeled forces themselves. In general, if one is interested \textcolor{black}{in} a  certain dynamical feature, \textcolor{black}{then} it \textcolor{black}{must} be explicitly modeled \textcolor{black}{in such a way that} one or more dedicated solve-for parameters are estimated. Subsequently, the resulting covariance matrix \textcolor{black}{can be examined to identify}   the correlations between  \textcolor{black}{various} parameters.
  Clearly, the magnitude of  post-fit range residuals can only be \textcolor{black}{greater} than, or \textcolor{black}{as large as} the range measurement precision. \textcolor{black}{Perfect models} and/or \textcolor{black}{total removal of all  effects that have not been modeled}  would \textcolor{black}{provide} residuals as large as the measurement precision.

\begin{figure}
\onefigure[width=8 cm]{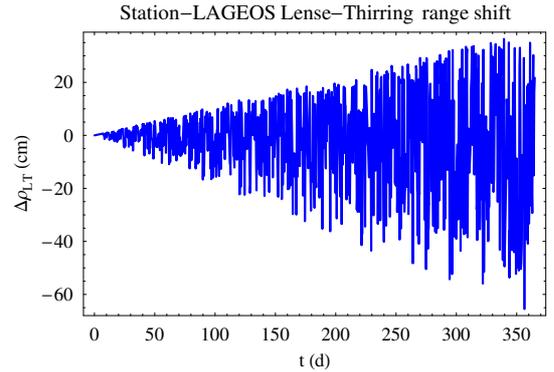}
\caption{Numerically integrated Lense-Thirring station-satellite range perturbation $\Delta\rho_{\rm LT}$ for LAGEOS over $\Delta T=1$ yr. Its variance is $18.0$ cm. \textcolor{black}{We choose the ITRF2000 coordinates of the GRAZ station, from \protect{\cite{cinesi}}: cut-off elevation angle of 20 deg.}}
\label{figura2}
\end{figure}

\begin{figure}
\onefigure[width=8 cm]{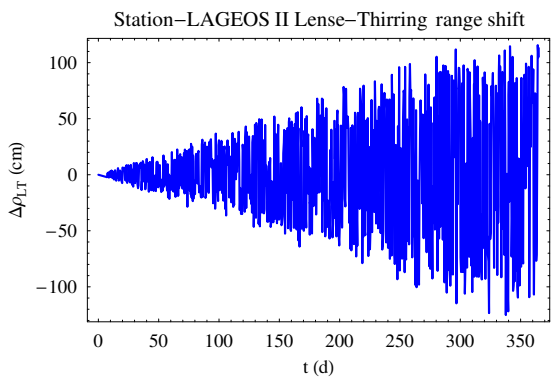}
\caption{Numerically integrated Lense-Thirring station-satellite range perturbation $\Delta\rho_{\rm LT}$ for LAGEOS II over $\Delta T=1$ yr. Its variance is $46.1$ cm. \textcolor{black}{We choose the ITRF2000 coordinates of the GRAZ station, from \protect{\cite{cinesi}}: cut-off elevation angle of 20 deg.}}
\label{figura3}
\end{figure}

\begin{figure}
\onefigure[width=8 cm]{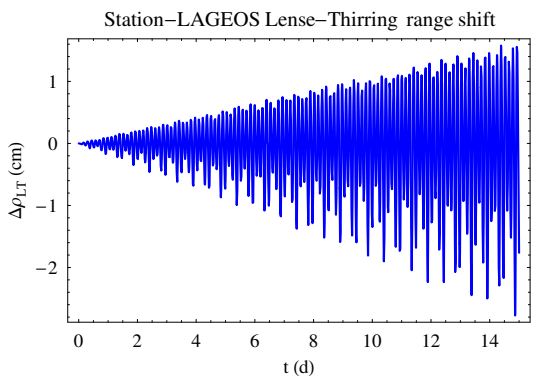}
\caption{Numerically integrated Lense-Thirring station-satellite range perturbation $\Delta\rho_{\rm LT}$ for LAGEOS over $\Delta T=15$ d. Its variance is $0.8$ cm. \textcolor{black}{We choose the ITRF2000 coordinates of the GRAZ station, from \protect{\cite{cinesi}}: cut-off elevation angle of 20 deg.}}
\label{figura4}
\end{figure}

\begin{figure}
\onefigure[width=8 cm]{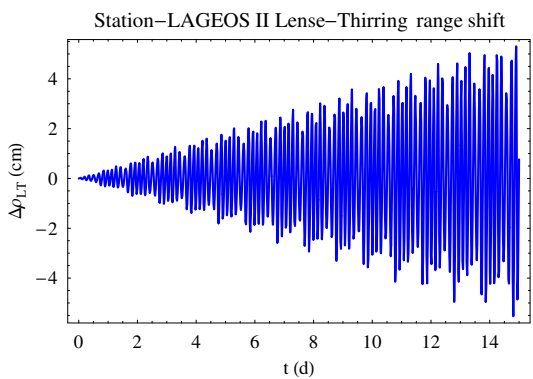}
\caption{Numerically integrated Lense-Thirring station-satellite range perturbation $\Delta\rho_{\rm LT}$ for LAGEOS II over $\Delta T=15$ d. Its variance is $1.9$ cm. \textcolor{black}{We choose the ITRF2000 coordinates of the GRAZ station, from \protect{\cite{cinesi}}: cut-off elevation angle of 20 deg.}}
\label{figura5}
\end{figure}

  Extending such considerations to the frame-dragging tests \textcolor{black}{made} so far with the LAGEOS satellites, it must be remarked that, actually, the Lense-Thirring force was never modeled, so that it should be considered \textcolor{black}{in the same way} as a source of systematic error impacting, in case, the post-fit range residuals to a certain level. No dedicated solve-for parameters were ever estimated; thus, the gravitomagnetic signature might have been partly or totally absorbed in the estimation of the several other parameters in the data reduction process, and partially or totally removed from the range signature.
  If frame-dragging fully impacted the ranges as predicted by general relativity, \textcolor{black}{there should be} time series of post-fit range residuals \textcolor{black}{with} the characteristic \textcolor{black}{signature} of the gravitomagnetic force itself. \textcolor{black}{See} Figure \ref{figura2} and Figure \ref{figura3} displaying the \textcolor{black}{numerically produced} nominal Lense-Thirring effect on the station-satellite range for LAGEOS and LAGEOS II over a time interval of $\Delta T=1$ yr. On the \textcolor{black}{other hand}, the same \textcolor{black}{set of data should be analyzed by explicitly modeling}  the Lense-Thirring effect in order to check if statistically significant differences with respect to the previous case would occur. \textcolor{black}{This} would be a crucial \textcolor{black}{test} of the \textcolor{black}{ability to actually measure terrestrial gravitomagentism} by means of the LAGEOS and LAGEOS II SLR data. In fact, after more than 15 years since the first tests, such \virg{gravitomagnetic} post-fit range residuals were never shown so far. It should be noticed that there is a contradiction between  claiming sub-cm post-fit range-residuals, obtained without modeling frame-dragging, and  figs. \ref{figura2} and \ref{figura3}
  displaying signatures with RMS variances as large as $18.0$ cm and $46.1$ cm, respectively.  Indeed, one should assume either that  the gravitomagnetic signal, not modeled,  was almost entirely removed or that it was almost canceled by the superposition of other unmodeled/mismodeled competing dynamical effects. After all, such a removal would not be \textcolor{black}{implausible} since, as shown by fig. \ref{figura4} and fig. \ref{figura5}, the nominal size of the Lense-Thirring range perturbation is just \textcolor{black}{at the level of cm on a timescale of} $\Delta t=15$ d.
\textcolor{black}{It is not clear, however,} why all the other  effects \textcolor{black}{not modeled at all, or poorly modeled,}  should   be exactly removed, \textcolor{black}{or should cancel each other} leaving just the completely unmodeled Lense-Thirring signal, \textcolor{black}{which is precisely what one expects to find in the data}. It is much more plausible that it is somewhat absorbed in some of the estimated parameters and removed from the residual signal to a certain extent.
\textcolor{black}{Somebody may  argue that the removal of the Lense-Thirring signature \textcolor{black}{can occur} only if certain once-per-revolution empirical cross-track accelerations were estimated. \textcolor{black}{First of all,} it should be explicitly proven that they were actually not estimated in the  dedicated LAGEOS data reductions. \textcolor{black}{More} importantly, it is impossible to a-priori decide in which of the estimated parameters the cancelation would actually \textcolor{black}{occur}. \textcolor{black}{Suffice} it to say that  in much more \virg{clean} scenarios like planetary astronomy, not plagued by the host of disturbances and non-gravitational effects of satellite geodesy, it is common practice to explicitly model the effects one is interested in and solve for one or more dedicated parameters just to avoid the risk that they may be partially or totally absorbed in the estimation of the initial state vectors. Interestingly, this has  been done \textcolor{black}{recently} \cite{pio1,pio2,pio3,pio4} even for \textcolor{black}{hypothetical} forces \textcolor{black}{that, as} the Pioneer Anomaly, if they \textcolor{black}{really existed in Nature} would have caused signatures much \textcolor{black}{greater} than the accuracy of the observations themselves.}

  We remark that the \textcolor{black}{LAGEOS-based} tests are likely plagued by another source of intrinsic a-priori imprinting of general relativity itself \textcolor{black}{in addition to}  those already pointed out \cite{review}. Indeed, they always made use of a reference system whose materialization heavily relies upon SLR data, among which those from LAGEOS and LAGEOS II themselves play a fundamental role.

The considerations exposed here are, in principle, valid also for other performed or proposed tests of general relativity with the LAGEOS satellites \cite{luc3}, and also for those which should be implemented in the near future with the existing LAGEOS and LAGEOS II, and with the new LARES satellite \cite{lares}, to be launched in \textcolor{black}{late} 2011  \textcolor{black}{with a VEGA rocket}.
\section{Conclusions}
In conclusion, \textcolor{black}{we can entertain} reasonable doubts \textcolor{black}{as to} what it was actually seen in the tests with  LAGEOS and LAGEOS II \textcolor{black}{made so far}, and \textcolor{black}{what has been}  passed of as frame-dragging in them. Only the use of a completely different \textcolor{black}{approach}, more \textcolor{black}{related} to \textcolor{black}{quantities that are actually measured}, \textcolor{black}{could afford to talk} about of  clear \textcolor{black}{and unambiguous} tests of \textcolor{black}{this} subtle effect. Frame-dragging should be explicitly modeled and solved-for in the LAGEOS and LAGEOS II data reduction process; post-fit range residuals produced with and without a model for the Lense-Thirring effect should be displayed and analyzed; a different materialization of the reference system used so far, mostly based on the observations of LAGEOS and LAGEOS II  themselves, should be adopted; \textcolor{black}{it would be preferable that GR is explicitly modeled} and solved-for in future dedicated global gravity field solutions combining data from several satellites.
Otherwise, \textcolor{black}{they} should \textcolor{black}{make clear}  why \textcolor{black}{they do not} implement the \textcolor{black}{strategy advocated here} which, after all, is standard practice in all branches of geodetic and astronomical studies.

 Moreover, even \textcolor{black}{accepting} the strategy \textcolor{black}{followed} so far, \textcolor{black}{the} unavoidable uncertainties in our knowledge of the  Earth's rotation axis  \textcolor{black}{affect the} necessarily imperfect calculation of the \textcolor{black}{theoretical} coefficient $c_1$ entering the linear combination of the nodes of LAGEOS and LAGEOS II. It \textcolor{black}{does} not allow \textcolor{black}{to obtain} an exact \textcolor{black}{cancelation} of the aliasing bias due to the first even zonal harmonic $J_2$ of the geopotential which, instead, would still be present at a $\simeq 20\%$ level of the Lense-Thirring signal. Let us recall that a further $10-20\%$ alias comes from the uncertainty in $c_1$ due to the errors in the satellites' orbital parameters $a$ and $I$.

 Thus, more work is still \textcolor{black}{needed} to really consider the LAGEOS-based attempt as  \textcolor{black}{a robust complement} of the GP-B mission from the point of view of reliability, trustability and methodology.  \textcolor{black}{Although} the LAGEOS-based tests \textcolor{black}{had} measured something \textcolor{black}{that really relates to} the Lense-Thirring effect, \textcolor{black}{their overall} uncertainty  \textcolor{black}{will probably make them less accurate than the} GP-B experiment. \textcolor{black}{Anyway}, independent analyses of the  data of the \textcolor{black}{Stanford team} by different groups are certainly required.

\acknowledgments
I thank L. Combrinck for having provided me with the state vectors of both the LAGEOS spacecrafts, and for interesting discussions.

\end{document}